\documentclass[12pt]{article}

\usepackage{amsmath}
\begin{document}

\title{Paraxial theory of sum-frequency generation by side-walk alignment and phase-matching in uniaxial crystals}
\author{Shahen Hacyan}

\renewcommand{\theequation}{\arabic{section}.\arabic{equation}}

\maketitle
\begin{center}

{\it  Instituto de F\'{\i}sica,} {\it Universidad Nacional Aut\'onoma
de M\'exico,}

{\it Apdo. Postal 20-364, M\'exico D. F. 01000, Mexico.}

e-mail: hacyan@fisica.unam.mx

\end{center}
\vskip0.5cm

\begin{abstract}
A formalism in terms of Hertz potentials is presented describing sum-frequency generation in a uniaxial non-linear
crystal. A scheme is proposed consisting in aligning the side-walk propagation of extraordinary waves in
combination with phase-matching. Simplified paraxial equations describing this situation are obtained. Particular
attention is paid to the generation of second harmonics.
\end{abstract}



\newpage

\section{Introduction}

Sum-frequency and difference-frequency generations are important processes in nonlinear optics (see, e.g., Ref.
\cite{boy,new}). The aim of the present article is to develop a formalism describing such nonlinear optical
processes in terms of Hertz potentials. The first part of the article is devoted to a derivation of the equations
describing the generation of higher frequencies  in a uniaxial nonlinear (up to second order) crystal in the
paraxial approximation. Particular attention is paid to the fact that, for extraordinary waves, the wave vectors
do not coincide with the group velocity vector. This usually originates some difficulties in optical experiments
and must be taken into account carefully \cite{dan}. In the second part of the article, a possible scheme is
proposed for generating higher frequencies using precisely this side-walk propagation. The idea is to combine
phase matching, which involves wave vectors, with group velocity vectors. More specifically, the scheme consists
in combining both ordinary and extraordinary beams in such a way that all three waves involved in the process are
aligned in the same direction. The conditions to be fulfilled by the crystal parameters for this particular
configuration are given explicitly. An analytic solution is also obtained for particular values of the amplitudes
of the initial waves.

The organization of this article is as follows. In section 2, a general formulation of the problem in terms of
Hertz potentials is worked out following Nisbet's original treatment \cite{nisbet}. The results are applied in
section 3 to the problem of sum-frequency generation. The possible alignment of the three group velocities
involved in the scheme is studied in section 4. The evolution equations are presented in section 5, together with
a particular analytic solution. Finally, the particular case of second harmonic generation is considered in
section 6.

\section{Propagation in a birefringent medium.}

Consider an anisotropic medium described by electric and magnetic field vectors, ${\bf E} $ and ${\bf B}$, and
electric displacement vector $ {\bf D}$. The Maxwell equations in the absence of free charges and currents (with
magnetic permeability $\mu=1$ and setting $c=1$) are
\begin{equation}
\nabla \cdot {\bf B} = 0~~, \quad \quad \nabla \times {\bf E} +
\frac{\partial {\bf B}}{\partial t}  =0~,
\end{equation}
\begin{equation}
\nabla \cdot {\bf D} = 0~~, \quad \quad \nabla \times {\bf B} -
\frac{\partial {\bf D }}{\partial t}  =0~.
\end{equation}

The effect of the material medium can be described by a polarization
vector ${\bf P}$ such that
$$
{\bf D} = {\bf E}+ 4\pi{\bf P},
$$
the linear part being given in terms of the dyad
$$
\widehat{\epsilon} = \epsilon_{\bot} {\bf 1} + \Delta \epsilon~{\bf s}~{\bf
s},
$$
where ${\bf s}$ is the optical axis of symmetry of the medium, and $\epsilon_{\bot}$ and $\epsilon_{\parallel}=\epsilon_{\bot} +
\Delta \epsilon$ are the permeability perpendicular and parallel to this symmetry axis respectively. Then
$$
{\bf D}(\omega, {\bf r}) = \widehat{\epsilon} (\omega) ~{\bf
E}(\omega, {\bf r}) + 4\pi {\bf P}^{NL}(\omega, {\bf r})
$$
where ${\bf P}^{NL}$ is the non-linear contribution to the polarization vector. Here and in the following, Fourier
transforms with respect to time of all quantities will be used.

Following Nisbet\cite{nisbet}, the electromagnetic field can be described by two scalar Hertz potentials, to be
called $\Psi_O$ and $\Psi_E$ in the present paper, and two additional scalar potentials, to be called $U_O$ and
$U_E$. These potentials satisfy the equations
\begin{equation}
\nabla_{\bot} U_E + {\bf s} \times \nabla U_O = 4\pi {\bf
P_{\bot}}^{NL} ~,\label{U}
\end{equation}
\begin{equation}
  \frac{1}{\epsilon_{\bot}
(\omega) }\nabla \cdot \widehat{\epsilon}(\omega) \cdot \nabla ~ \Psi_E + \epsilon_{\parallel}(\omega) \omega^2
\Psi_E - \frac{\epsilon_{\parallel}(\omega)}{\epsilon_{\bot}(\omega)} ~ {\bf s}  \cdot \nabla U_E = -4\pi
P^{NL}_{\parallel}~,
\end{equation}
\begin{equation}
 \nabla^2 \Psi_O + \epsilon_{\bot}(\omega)\omega^2  \Psi_O +  i\omega U_O =0~.
\end{equation}
In these formulas, $\nabla_{\bot}$ is the gradient operator in the
plane perpendicular to ${\bf s}$. Eq. (\ref{U}) implies
\begin{equation}
\nabla_{\bot}^2 U_E = 4 \pi ~\nabla_{\bot} \cdot {\bf
P_{\bot}}^{NL}~,
\end{equation}
\begin{equation}
\nabla_{\bot}^2 U_O = 4 \pi ~{\bf s} \cdot (\nabla \times {\bf
P}^{NL})~,
\end{equation}
which permits to decouple the potentials $\Psi_O$ and $U_O$ from $\Psi_E$ and $U_E$.

As shown in a previous article (Hacyan and J\'auregui\cite{hj09}),
the advantage of this formulation is that $\Psi_O$ and $\Psi_E$
correspond to the potentials for the ordinary and extraordinary waves
respectively. The electromagnetic field is given by
\begin{equation}
{\bf E} =  ~-i\omega {\bf s} \times \nabla \Psi_O +
\frac{1}{\epsilon_{\bot}} \nabla ({\bf s} \cdot \nabla \Psi_E) + \omega^2
\Psi_E ~ {\bf s} - \frac{1}{\epsilon_{\bot}} \nabla U_E \label{Etc}
\end{equation}
and
\begin{equation}
{\bf B} =   \nabla \times [\nabla \times ({\Psi}_O {\bf s})] -i\omega
\nabla \times (\Psi_E {\bf s}) \label{Btc}~.
\end{equation}
From these formulas, the two fundamental modes can be identified: the
ordinary wave with ${\bf s} \cdot {\bf E}_O=0$ and the extraordinary
wave with ${\bf s} \cdot {\bf B}_E=0$. The case ${\bf P}^{NL}=0$ with
$U_{O,E}=0$ corresponds to the linear limit considered in
\cite{hj09}.

The non-linear polarization vector is usually defined as
\begin{equation}
P^{NL}_a (\omega, {\bf r}) = \int d\omega_1 \int d\omega_2 ~\delta
(\omega - \omega_1 -\omega_2) \chi_{abc} (\omega_1 , \omega_2) E_b
(\omega_1, {\bf r}) E_c (\omega_2, {\bf r})
\end{equation}
in the quadratic approximation, where $\chi_{abc} (\omega_1 ,
\omega_2)$ is the (Fourier transformed) second-order susceptibility
tensor (assumed to be homogeneous).

A particularly important case is the one in which there is a discrete
set of well defined frequencies $\omega_i$, such that
$$
E_a (\omega, {\bf r}) = \sum_i \delta (\omega -\omega_i) E^{(i)}_a ({\bf r})~.
$$
Then the basic equations take the form
\begin{equation}
\nabla_{\bot} U_E^{(i)}({\bf r}) + {\bf s} \times \nabla
U_O^{(i)}({\bf r}) = 4\pi {\bf P}_{\bot}^{(i)}(\omega_j ,\omega_k
,{\bf r})~, \label{Uom}
\end{equation}
and
\begin{equation}
-\Big[\epsilon_{\parallel}(\omega_i) \omega_i^2  +
\frac{1}{\epsilon_{\bot}(\omega_i)} \nabla \cdot
\widehat{\epsilon}(\omega_i) \cdot \nabla ~ \Big] \Psi_E^{(i)}({\bf
r}) + \frac{\epsilon_{\parallel}(\omega_i)}{\epsilon_{\bot} (\omega_i)} ~
{\bf s}  \cdot \nabla U_E^{(i)}({\bf r}) = 4\pi
P^{(i)}_{\parallel}(\omega_j ,\omega_k ,{\bf r})
\end{equation}
for extraordinary waves and
\begin{equation}
\Big[\epsilon_{\bot} (\omega_i) \omega_i^2 +  \nabla^2 \Big]
\Psi_O^{(i)}({\bf r}) + i \omega_i U_O^{(i)}({\bf r}) =0
\end{equation}
for ordinary waves,  where (setting $\omega_i = \omega_j + \omega_k$)
\begin{equation}
 P_a^{(i)}(\omega_i =\omega_j +\omega_k ,{\bf r}) =
 \chi_{abc} (\omega_i =\omega_j +\omega_k) E_b^{(j)} ({\bf r})
E_c^{(k)} ({\bf r})
\end{equation}
and
\begin{equation}
 P_a^{(j)}(\omega_j = \omega_i -\omega_k ,{\bf r}) =
 \chi_{abc} (\omega_j = \omega_i -\omega_k ) E_b^{(i)} ({\bf r})
E_c^{(k)*} ({\bf r})~.
\end{equation}

\section{Sum-frequency generation}
\setcounter{equation}{0}

Consider a typical problem of sum-frequency generation. Suppose an ordinary and an extraordinary waves, of frequencies
$\omega_1$ and $\omega_2$ respectively, combine inside the crystal to generate an extraordinary wave of frequency
$\omega_3 = \omega_1 + \omega_2$. Let $\psi_i$ be the Hertz potential corresponding to frequencies $\omega_i$, and
$U_i$ the associated auxiliary potentials. Accordingly the basic equations take the form:
\begin{equation}
\Big[\omega_1^2 \epsilon_{\bot}(\omega_1)   + \nabla^2 ~ \Big] \Psi_1({\bf
r}) = - i \omega_1  U_1({\bf r}) ~,  \label{sfO}
\end{equation}
\begin{eqnarray}
\Big[\omega_2^2 \epsilon_{\bot}(\omega_2) \epsilon_{\parallel}(\omega_2)   &+&
\nabla \cdot
\widehat{\epsilon}(\omega_2) \cdot \nabla ~ \Big] \Psi_2({\bf r}) \\ \nonumber
&=& \epsilon_{\parallel}(\omega_2) ~ {\bf s} \cdot \nabla U_2({\bf r}) - 4\pi \epsilon_{\bot} (\omega_2)
P^2_{\parallel}(\omega_2 = \omega_3 -\omega_1 ,{\bf r})~,\label{sfE2}
\end{eqnarray}
\begin{eqnarray}
\Big[\omega_3^2 \epsilon_{\bot} (\omega_3) \epsilon_{\parallel}(\omega_3)   &+&
\nabla \cdot
\widehat{\epsilon}(\omega_3) \cdot \nabla ~ \Big] \Psi_3({\bf r})  \\ \nonumber
&=& \epsilon_{\parallel}(\omega_3) ~ {\bf
s} \cdot \nabla U_3({\bf r}) - 4\pi \epsilon_{\bot} (\omega_3) P^3_{\parallel}(\omega_3 =
\omega_1 +\omega_2 ,{\bf r})~,\label{sfE3}
\end{eqnarray}
and
\begin{equation}
\nabla_{\bot}^2 U_1 = 4 \pi ~{\bf s} \cdot [\nabla \times {\bf
P}_{\bot}^1 (\omega_1 = \omega_3 -\omega_2 ,{\bf r})~]~,
\end{equation}
\begin{equation}
\nabla_{\bot}^2 U_2 = 4 \pi ~\nabla_{\bot} \cdot {\bf P}^2 (\omega_2 = \omega_3 - \omega_1 ,{\bf
r})~~,\label{sfP2}
\end{equation}
\begin{equation}
\nabla_{\bot}^2 U_3 = 4 \pi ~\nabla_{\bot} \cdot {\bf P}^3 (\omega_3
= \omega_1  + \omega_2 ,{\bf r})~~,\label{sfP3}
\end{equation}
where
\begin{eqnarray}
 P_a^1(\omega_1 = \omega_3 -\omega_2  ,{\bf r}) &=&
 \chi_{abc} (\omega_1 = \omega_3 -\omega_2 ) E_b^3({\bf r})
[E_c^2 ({\bf r})]^* \\ \nonumber
 P_a^2(\omega_2 = \omega_3 -\omega_1  ,{\bf r}) &=&
 \chi_{abc} (\omega_2 = \omega_3 -\omega_1 ) E_b^3({\bf r})
[E_c^1 ({\bf r})]^* \\ \nonumber
 P_a^3(\omega_3
= \omega_1  + \omega_2 ,{\bf r}) &=&
 \chi_{abc} (\omega_3
= \omega_1  + \omega_2) E_b^1({\bf r}) E_c^2 ({\bf r}).\\ \nonumber
\end{eqnarray}

As a next step, let us assume that the potentials have the form
\begin{equation}
\psi_i = A_i ({\bf r}) e^{i {\bf k}_i \cdot {\bf r}}
\end{equation}
where
\begin{equation}
{\bf k}_1^2 = \epsilon_{\bot} (\omega_1) ~\omega_1^2
\end{equation}
and
\begin{equation}
{\bf k}_j \cdot \widehat{\epsilon}(\omega_j) \cdot {\bf k}_j =
\epsilon_{\bot} (\omega_j) {\bf k}_{j\bot}^2 +\epsilon_{\parallel}(\omega_j)
{\bf k}_{j\parallel}^2 =
\epsilon_{\bot}(\omega_j)\epsilon_{\parallel}(\omega_j) ~\omega_j^2~,
\label{kk}
\end{equation}
for $j=2,3$, and also
\begin{equation}
U_1 = u_1({\bf r}) e ^{i({\bf k}_3 - {\bf k}_2) \cdot {\bf r}}~,
\end{equation}
\begin{equation}
U_2 = u_2({\bf r}) e ^{i({\bf k}_3 - {\bf k}_1) \cdot {\bf r}}~,
\end{equation}
\begin{equation}
U_3 = u_3({\bf r}) e ^{i({\bf k}_1 +  {\bf k}_2) \cdot {\bf r}}~.
\end{equation}
In the above equations, $A_i({\bf r})$ and $u_i({\bf r})$ are slowly varying functions
of ${\bf r}$.

 Within this same approximation:
\begin{equation}
{\bf E}_1 \simeq \omega_1 ~{\bf s} \times {\bf k}_1 A_1({\bf r})
e^{i{\bf k}_1 \cdot {\bf r}}
\end{equation}
and
\begin{equation}
{\bf E}_j \simeq  \Big[  \omega_j^2 {\bf s} -\frac{1}{\epsilon_{\bot}
(\omega_j)} ({\bf s} \cdot {\bf k}_j) {\bf k}_j  \Big] A_j({\bf r})
~e ^{i{\bf k}_j \cdot {\bf r}}  - \frac{1}{\epsilon_{\bot} (\omega_j)}
\nabla U_j~.
\end{equation}
for $j=1,2$. The last term in the above equation is quadratic in the
electromagnetic  field; to be consistent, it must be neglected when
evaluating the polarization vector up to second order. Accordingly:
\begin{eqnarray} {\bf P}_1(\omega_1
= \omega_3  - \omega_2, {\bf r}) &=& {\bf p}_1 ~ A_3({\bf r})
A_2^*({\bf r}) e ^{ i({\bf k}_3 - {\bf k}_2 )\cdot {\bf r}} ~, \\ \nonumber
 {\bf P}_2(\omega_2
= \omega_3  - \omega_1, {\bf r}) &=& {\bf p}_2 ~ A_3({\bf r})
A_1^*({\bf r}) e ^{ i({\bf k}_3 - {\bf k}_1 )\cdot {\bf r}} ~, \\ \nonumber
{\bf P}_3(\omega_3 = \omega_1  + \omega_2, {\bf r}) &=& {\bf p}_3 ~ A_1 ({\bf r})A_2 ({\bf r}) e ^{ i({\bf k}_1 +
{\bf k}_2) \cdot {\bf r}} ~,\\ \nonumber
\end{eqnarray}
where the vectors ${\bf p}_i$ are given in terms of $\chi_{abc}$,
${\bf s}$ and ${\bf k}_i$ as
\begin{eqnarray}
p_{1,a} &=& \chi_{abc} (\omega_1 = \omega_3 - \omega_2)~ e_{3,b}~ e_{2,c}~,\label{ppp} \\ \nonumber p_{2,a} &=&
\chi_{abc} (\omega_2 = \omega_3 - \omega_1)~ e_{3,b}~ e_{1,c}~, \\ \nonumber p_{3,a} &=& \chi_{abc} (\omega_3 =
\omega_1 + \omega_2)~ e_{1,b}~ e_{2,c}~,
\end{eqnarray}
with
\begin{equation}
{\bf e}_1 = \omega_1 {\bf s}\times{\bf k}_1~,
\end{equation}
\begin{equation}
{\bf e}_j = \omega_j^2 {\bf s} - \frac{1}{\epsilon_{\bot} (\omega_j)} ({\bf
s} \cdot {\bf k}_j) {\bf k}_j~,
\end{equation}
for $j=2,3$.

The basic equations (\ref{sfO}) to (\ref{sfP3}) now take the form
\begin{equation}
\nabla^2 A_1 + 2 i {\bf k}_1 \cdot \nabla A_1 = -i\omega_1 u_1 e^{-i
\Delta {\bf k} \cdot {\bf r}},
\end{equation}
\begin{eqnarray}
&& \nabla \cdot \hat{\epsilon}(\omega_2) \cdot  \nabla A_2 +   2 i {\bf k}_2 \cdot \hat{\epsilon}(\omega_2) \cdot \nabla A_2 \\
\nonumber &=& \Big[ ~i \epsilon_{\parallel}(\omega_2 )~ {\bf s} \cdot ({\bf k}_3 - {\bf k}_1) u_2 - 4 \pi
\epsilon_{\bot}(\omega_2) p_{2\parallel} A_1^* A_3 \Big] e^{-i\Delta {\bf k} \cdot {\bf r}}~,
\end{eqnarray}
\begin{eqnarray}
&& \nabla \cdot \hat{\epsilon}(\omega_3) \cdot  \nabla A_3 +   2 i {\bf k}_3 \cdot \hat{\epsilon}(\omega_3) \cdot \nabla A_3 \\
\nonumber &=& \Big[ ~i \epsilon_{\parallel}(\omega_3 )~ {\bf s} \cdot
({\bf k}_1 + {\bf k}_2) u_3 - 4 \pi  \epsilon_{\bot}(\omega_3)
p_{3\parallel} A_1 A_2 \Big] e^{i\Delta {\bf k} \cdot {\bf r}}~,
\end{eqnarray}
where $\Delta {\bf k}={\bf k}_1 + {\bf k}_2 - {\bf k}_3$, and
\begin{equation}
 -({\bf k}_{3\bot} - {\bf k}_{2\bot})^2 u_1 =  4 \pi i ~[{\bf s} \times ({\bf k}_3
  - {\bf k}_2) ] \cdot {\bf p}_{1 \bot} ~A_2^* A_3
\end{equation}
\begin{equation}
 -({\bf k}_{3\bot} - {\bf k}_{1\bot})^2 u_2 =  4\pi i ~ ({\bf k}_{3\bot}
- {\bf k}_{1\bot}) \cdot {\bf p}_{2 \bot} ~A_1^* A_3~,
\end{equation}
and
\begin{equation}
-({\bf k}_{1\bot} + {\bf k}_{2\bot})^2 u_3= 4\pi i ~ ({\bf k}_{1\bot}
+ {\bf k}_{2\bot}) \cdot {\bf p}_{3 \bot} ~A_1 A_2~,
\end{equation}
within the same approximation (that is, keeping only terms of order
${\bf k}^2$).

\subsection{Phase matching}

The phase matching condition is ${\bf k}_3 = {\bf k}_1 + {\bf k}_2$,
in which case the above equations somewhat simplify:
\begin{equation}
\nabla^2 A_1 + 2i  {\bf k}_1 \cdot \nabla A_1 = -i\omega_1 u_1
,\label{ph-m1}
\end{equation}
\begin{equation}
\nabla \cdot \hat{\epsilon} (\omega_2) \cdot \nabla A_2 + 2 i  {\bf
k}_2 \cdot \hat{\epsilon}(\omega_2) \cdot \nabla A_2 =
 ~i \epsilon_{\parallel}(\omega_2 )~ {\bf s} \cdot {\bf k}_2 u_2
- 4 \pi  \epsilon_{\bot}(\omega_2) p_{2\parallel} A_1^* A_3 ~,\label{ph-m2}
\end{equation}
\begin{equation}
\nabla \cdot \hat{\epsilon} (\omega_3) \cdot \nabla A_3 + 2 i  {\bf
k}_3 \cdot \hat{\epsilon}(\omega_3) \cdot \nabla A_3 =
 ~i \epsilon_{\parallel}(\omega_3 )~ {\bf s} \cdot {\bf k}_3 u_3
- 4 \pi  \epsilon_{\bot}(\omega_3) p_{3\parallel} A_1 A_2~, \label{ph-m3}
\end{equation}
and
\begin{eqnarray}
 -({\bf k}_{1\bot})^2 u_1 &=&  4 \pi i ~({\bf s} \times {\bf k}_1) \cdot
 {\bf p}_1  A_2^* A_3~, \\ \nonumber
 -({\bf k}_{2\bot})^2 u_2 &=&  4 \pi i ~{\bf k}_2 \cdot {\bf p}_{2\bot} A_1^* A_3 ~,\\ \nonumber
-({\bf k}_{3\bot})^2 u_3&=& 4\pi i ~ {\bf k}_3 \cdot {\bf p}_{3\bot}
A_1 A_2~.\\ \nonumber
\end{eqnarray}

Now, if no absorption is present, we have the following general relations:
$$
\chi^*_{abc} (\omega_3 = \omega_1 + \omega_2)= \chi_{bca} (\omega_1 = \omega_3 - \omega_2)=\chi_{cab} (\omega_2 =
\omega_3 - \omega_1)~
$$
(Kleinman \cite{kle}). Eqs. (\ref{ppp}) then imply
\begin{equation}
{\bf e}_1  \cdot {\bf p}_1 = {\bf e}_2  \cdot {\bf p}_2= {\bf e}_3
\cdot {\bf p}_3^* \equiv -C~.
\end{equation}

With this last condition, it follows after some lengthy but
straightforward algebra [taking relations (\ref{kk}) into account]
that the basic equations (\ref{ph-m1}-\ref{ph-m3}) take the form
\begin{equation}
\nabla^2 A_1 + 2 i {\bf k}_1 \cdot \nabla A_1 = ~\frac{4\pi }{|{\bf
k}_{1\bot}|^2}~C A_2^* A_3  ~ ,\label{331}
\end{equation}
\begin{equation}
\nabla \cdot \hat{\epsilon}(\omega_2) \cdot  \nabla A_2+  2 i {\bf
k}_2 \cdot \hat{\epsilon}(\omega_2) \cdot \nabla A_2 = \frac{4\pi
}{|{\bf k}_{2\bot}|^2}
 ~\epsilon_{\parallel}(\omega_2 )\epsilon(\omega_2 )~ C A_1^* A_3~  ,
\end{equation}
\begin{equation}
\nabla \cdot \hat{\epsilon}(\omega_3 ) \cdot  \nabla A_3+  2 i {\bf
k}_3 \cdot \hat{\epsilon}(\omega_3) \cdot \nabla A_3 = \frac{4\pi
}{|{\bf k}_{3\bot}|^2}
 ~\epsilon_{\parallel}(\omega_3 )\epsilon(\omega_3 )~ C^* A_1 A_2~.\label{333}
\end{equation}

\section{Side-walk alignment}
\setcounter{equation}{0}

Suppose we want to generate a frequency $ \omega_3 = \omega_1 +\omega_2$. Since rays 2 and 3 are extraordinary,
they do not propagate along ${\bf k}_2$ and ${\bf k}_3$, but rather along the directions $ \hat{\epsilon}(2) \cdot
{\bf k}_2 $ and $ \hat{\epsilon}(3) \cdot {\bf k}_3$ respectively, due to the side-walk effect (see e.g., Ref.
\cite{hac}) . It it then possible to choose the directions of propagations in such a way that the three rays
propagates along the same direction, \emph{in addition} to the phase matching condition.
 This can be achieved setting
$$
{\bf k}_1 + {\bf k}_2 = {\bf k}_3~,
$$
$$
\widehat{\epsilon}(2) \cdot {\bf k}_2 = \epsilon_{\bot}(2){\bf k}_{2\bot} + \epsilon_{\parallel}(2) k_{2
\parallel } {\bf s} = \alpha {\bf k}_1
$$
\begin{equation}
\widehat{\epsilon}(3) \cdot {\bf k}_3 =\epsilon_{\bot}(3){\bf k}_{3\bot} + \epsilon_{\parallel}(2) k_{3\parallel }
{\bf s} = \beta {\bf k}_1\label{41}
\end{equation}
(in this section, we set $\epsilon (\omega_i) \rightarrow \epsilon(i)$ in order to lighten the notation). This
system of linear equations admits non-trivial solutions if the proportionality constants $\alpha$ and $\beta$ take
the values:
$$
\alpha = \frac{\epsilon_{\parallel}(2)\epsilon_{\bot}(2) \Delta \epsilon (3)}{D}
$$
\begin{equation}
\beta = \frac{\epsilon_{\parallel}(3)\epsilon_{\bot}(3) \Delta \epsilon(2)}{D}~,\label{42}
\end{equation}
where
\begin{equation}
D \equiv \epsilon_{\bot}(3) \Delta \epsilon(2)
-\epsilon_{\bot}(2) \Delta \epsilon(3).
\end{equation}
Furthermore, it is
evident that the three ray vectors ${\bf k}_i$ and the optical axis
${\bf s}$ must be in the same plane.

Dividing the wave vectors into components perpendicular and parallel
to ${\bf s}$, we have additionally the conditions
\begin{equation}
k_{1\bot}^2 +
k_{1\parallel}^2=\epsilon_{\bot}(1)~\omega_1^2~,\label{k1}
\end{equation}
\begin{equation}
\frac{k_{2\bot}^2}{\epsilon_{\parallel} (2)} +
\frac{k_{2\parallel}^2}{\epsilon_{\bot} (2)}
=~\omega_2^2~,\label{k2}
\end{equation}
\begin{equation}
\frac{k_{3\bot}^2}{\epsilon_{\parallel} (3)} +
\frac{k_{3\parallel}^2}{\epsilon_{\bot} (3)}
=~\omega_3^2~,\label{k3}
\end{equation}
and since
$$
k_{2\bot}=\frac{\alpha}{\epsilon (2)}k_{1\bot} ~,\quad
k_{2\parallel}=\frac{\alpha}{\epsilon_{\parallel}
(2)}k_{1\parallel}~,
$$
$$
k_{3\bot}=\frac{\beta}{\epsilon (3)}k_{1\bot} ~,\quad
k_{3\parallel}=\frac{\beta}{\epsilon_{\parallel}
(3)}k_{1\parallel}~,
$$
it follows from (\ref{k2}) that
\begin{equation}
\Delta \epsilon (2) k_{1\bot}^2 =  \Big(\frac{\omega_2D}{\Delta \epsilon
(3)}\Big)^2-\epsilon_{\bot}(1)\epsilon_{\bot}(2)\omega_1^2~,
\end{equation}
\begin{equation}
\Delta \epsilon (2) k_{1\parallel}^2 = - \Big(\frac{\omega_2 D}{\Delta \epsilon
(3)}\Big)^2+\epsilon_{\bot}(1)\epsilon_{\parallel}(2)\omega_1^2~.
\end{equation}
From (\ref{k3}) it also follows that
\begin{equation}
\Delta \epsilon (3) k_{1\bot}^2 = \Big(\frac{\omega_3 D}{\Delta \epsilon
(2)}\Big)^2-\epsilon_{\bot}(1)\epsilon_{\bot}(3)\omega_1^2~,
\end{equation}
\begin{equation}
\Delta \epsilon (3) k_{1\parallel}^2 = - \Big(\frac{\omega_3 D}{\Delta \epsilon
(2)}\Big)^2+\epsilon_{\bot}(1)\epsilon_{\parallel}(3)\omega_1^2~.
\end{equation}

Accordingly, the following relation is necessary for consistency:
\begin{equation}
 \epsilon (1) \omega_1^2 = D
\Big(  \frac{\omega_3^2}{\Delta \epsilon (2)}  -
\frac{\omega_2^2}{\Delta \epsilon (3)} \Big)~,\label{cons}
\end{equation}
besides, of course, $\omega_1 + \omega_2 = \omega_3$.

From the above formulas, it follows with some lengthy but straightforward algebra that the angle $\theta_1$
between ${\bf k}_1$ and ${\bf s}$ are given by the following equivalent formulas:
\begin{eqnarray}
\sin^2 \theta_1 = \frac{k_{1 \bot}^2}{k_{1 \parallel^2}} &=& \frac{1}{\epsilon_{\bot}(1) \Delta \epsilon(2)}
\Big(\frac{\omega_2 D}{\omega_1\Delta \epsilon(3)}
\Big)^2 - \frac{\epsilon_{\bot}(2)}{\Delta \epsilon(2)} \\
\nonumber &=&\frac{1}{\epsilon_{\bot}(1) \Delta \epsilon(3)} \Big(\frac{\omega_3 D}{\omega_1 \Delta \epsilon(2)}
\Big)^2 - \frac{\epsilon_{\bot}(3)}{\Delta \epsilon(3)} \\ \nonumber &=& \frac{[\omega_2 \Delta \epsilon
(2)]^2\epsilon_{\bot} (3) - [\omega_3 \Delta \epsilon (3)]^2\epsilon_{\bot} (2)}{\Delta \epsilon (2) \Delta
\epsilon (3) ~[\omega_3^2  \Delta \epsilon (3)- \omega_2^2 \Delta \epsilon (2)]}~.
\end{eqnarray}

The consistency conditions for these equations (since $1 >\sin^2 \theta > 0$) are
\begin{equation}
\Big[ \omega_3 \Delta \epsilon (3)\Big]^2 \epsilon_{\parallel} (2) >(<) \Big[ \omega_2 \Delta \epsilon (2)\Big]^2
\epsilon_{\parallel} (3)
\end{equation}
if $\omega_3^2 \Delta \epsilon (3) > (<) \omega_2^2 \Delta \epsilon (2)$.

\section{Evolution equations}
\setcounter{equation}{0}

Eqs. (\ref{331}-\ref{333}) simplify considerably under the assumption that the only relevant spatial variations
are along ${\bf k}_1$. Choosing the $z$ axis along that direction, it follows that
$$
\frac{d}{dz} A_1 = ~C_1  A_2^* A_3,
$$
$$
\frac{d}{dz} A_2 = ~C_2  A_1^* A_3 ,
$$
\begin{equation}
\frac{d}{dz} A_3 = ~-~ C_3^* A_1 A_2~,\label{CCC}
\end{equation}
where, using (\ref{41}) and (\ref{42}),
\begin{equation}
C_1 =\frac{4\pi}{2i k_1 |{\bf k}_{\bot 1}|^2} C \label{C1}
\end{equation}
\begin{equation}
C_2 =\frac{D^3}{\epsilon_{\parallel}^2(2)  \Delta \epsilon^3 (3)}C_1 \label{C2}
\end{equation}
\begin{equation}
C_3 =\frac{D^3}{\epsilon_{\parallel}^2(3)  \Delta \epsilon^3 (2)}C_1~. \label{C3}
\end{equation}
Notice that all three coefficients $C_1$, $C_2$ and $C_3$ are complex, but have the same phase.

From the above equations it follows that there is a conserved
quantity:
\begin{equation}
\frac{d}{dz} \Big( C_2 C_3^* |A_1|^2 + C_3 C_1^* |A_2|^2 - C_1
C_2^* |A_3|^2 \Big) =0~,
\end{equation}
which is the Manley-Rowe relation \cite{mr}.

A particular solution of Eqs. (\ref{CCC}) is
$$
A_1= a_1 e^{i\delta} ~ {\rm sech} (\gamma z)
$$
$$
A_2= a_2 e^{i\delta}~{\rm sech} (\gamma z)
$$
$$
A_3= a_3 e^{i\delta} \tanh (\gamma z)~,
$$
where $\delta$ is the common phase of $C_i$ and $a_i$ are real coefficients. Since the amplitudes of $A_1$ and
$A_2$ are given as initial conditions, the amplitude $A_3$ of the generated wave follows from the relation
\begin{equation}
a_3 = -|C_3| a_2 a_2 \gamma^{-1},
\end{equation}
with
\begin{equation}
\gamma^2 = |C_1|| C_3|a_2^2 = |C_2|| C_3|a_1^2.
\end{equation}
Thus the additional condition $|C_1|/|C_2|=a_1^2 /a_2^2$ must be fulfilled for the above analytic solution to be
valid. The ratio $|C_1| /|C_2|$ follows directly from Eqs. (\ref{C1}) and (\ref{C2}).

\section{Second harmonic generation}
\setcounter{equation}{0}

Let us consider as a further example the generation of second harmonics by non-linear effects. Usually, under
appropriate conditions, an ordinary wave of frequency $\omega$ gives rise to an extraordinary wave of frequency
$2\omega$. Accordingly, the process is described by the equations given above, with the following identification:
${\bf k}_1$ and ${\bf k}_2$ correspond to the ordinary and extraordinary rays respectively, both with frequency
$\omega$, and ${\bf k}_3$ corresponds to the extraordinary wave of frequency $2 \omega$, that is: $ \omega_1 =
\omega_2 \equiv \omega$ and $\omega_3 =2 \omega$.

In order to further lighten the notation, let us redefine $\epsilon_{\bot} (\omega) \equiv \epsilon$ and
$\epsilon_{\bot} (2\omega) \equiv \overline{\epsilon}$, and similarly for $\Delta \epsilon$ and $\epsilon_{\parallel}$.

Then, according to the consistency condition (\ref{cons}):
\begin{equation}
\epsilon = D \Big( \frac{4}{\Delta \epsilon} -\frac{1}{\Delta
\overline{\epsilon}} \Big),
\end{equation}
from where it follows, using the definition of $D$, that
\begin{equation}
\frac{\epsilon}{\overline{\epsilon}} = 1 - \Big[
\frac{\Delta \epsilon}{2\Delta \overline{\epsilon}} -1 \Big]^2 \label{62}
\end{equation}
and therefore
\begin{equation}
D =  \frac{\overline{\epsilon} \Delta \epsilon^2}{4\Delta \overline{\epsilon}}.
\end{equation}
It also follows from (\ref{62}) that
\begin{equation}
0 < \frac{\Delta \epsilon}{\Delta \overline{\epsilon}}< 4~.\label{cond}
\end{equation}
This inequality must be satisfied in order to have triple alignment of the velocity vectors.

Also
\begin{equation}
\sin^2 \theta_1 = \frac{\Delta \epsilon^2  \overline{\epsilon} - 4 \epsilon \Delta \overline{\epsilon}^2}{\Delta \epsilon \Delta \overline{\epsilon} (4 \Delta \overline{\epsilon}  - \Delta \epsilon)}.
\end{equation}
Thus, if the optical axis ${\bf s}$ makes an angle
$\phi$ with the unit normal vector  to the surface of the
crystal, then according to Snel's law,
\begin{equation}
\sin \iota = \sqrt{\epsilon} \sin (\theta_1 - \phi),
\end{equation}
where $\iota$ is the incidence angle to which the impinging ray must be directed in order to have a phase-matching
assisted by side-walk alignment. Equation (\ref{cond}) must be satisfied.

The evolution of the field is given by Eqs. (\ref{CCC}) with its coefficient given by
$$
C_2 = \frac{\overline{\epsilon}^3}{\epsilon^2_{\parallel}} \Big( \frac{\Delta \epsilon}{2 \Delta
\overline{\epsilon}} \Big)^6 C_1~,
$$
\begin{equation}
C_3= \frac{\overline{\epsilon}^3}{\overline{\epsilon}^2_{\parallel}} \Big( \frac{\Delta \epsilon}{2 \Delta
\overline{\epsilon}} \Big)^6 C_1~.
\end{equation}

\section{Concluding remarks}

The formalism presented in this paper can be applied to other processes, such as difference-frequency generation
and parametric down-conversion (to be considered in a forthcoming publication). As for the particular scheme of
side-walk alignment herein proposed, it is left as a proposal to find crystals with the appropriate parameters,
and to check its validity experimentally.

Work supported by PAPIIT-UNAM project IN101511.

\end{document}